MAGNETIC AND RESONANCE PROPERTIES OF THE COMPOUND
$(NH_3)_2(CH_2)_3CoCl_4$ — AN ANTIFERROMAGNET WITH THE DZYALOSHINSKII INTERACTION.

A.S. Cherny, K.G. Dergachev, M.I. Kobets, and E.N. Khatsko.

B. Verkin Institute for Low Temperature Physics and Engineering, National Academy of Sciences
of Ukraine, pr. Lenina 47, Kharkov 61103, Ukraine

E-mail: khatsko@ilt.kharkov.ua

Abstract

The static magnetic and dynamic properties of the compound $(NH_3)_2(CH_2)_3CoCl_4$ are investigated in the temperature interval 0.5–50 K. It is shown that at $T_N = 2.05$ K this compound undergoes a transition to a magnetically ordered antiferromagnet state. A distinctive feature of this compound is the presence of temperature hysteresis at the transition to the ordered state (corresponding to a first-order magnetic phase transition) and of a spontaneous magnetic moment along the $x$ axis. The paramagnetic Curie temperatures are determined. The frequency–field curves of the AFMR spectrum in the xy plane are investigated at a temperature below $T_N$. The main magnetic parameters of the biaxial AFM — the value of the low-frequency gap and the effective magnetic anisotropy field — are determined.

By the late 1970s a large number of new compounds manifesting properties close to one-dimensional and two-dimensional magnetic models had been found. Among them was the family of many-sublattice metalorganic perovskitelike compounds with the generalize chemical formula $[NH_3–(CH_2)_n–NH_3]MCl_4$ (n =1, 2, 3,...; M=$Mn^{2+}$, $Cu^{2+}$, $Fe^{2+}$, $Co^{2+}$), [1] in which the ratio of the interlayer exchange to the intralayer energy ($J_0/J$) reaches values of 10.8 . The structure of these compounds is formed by layers of iron-group ions separated by long organic molecules. The best studied [2, 3] magnetic crystals at the present time are compounds with $Mn^{2+}$ and $Cu^{2+}$. Substitution of some of the iron-group ions alters not only the magnetic and esonance properties but also the type of magnetic ordering. For example, compounds containing $Mn^{2+}$ have three-dimensional antiferromagnetic ordering, while in compounds containing $Cu^{2+}$ the intralayer interaction is ferromagnetic.

We synthesized the compound $(NH_3)_2(CH_2)_3CoCl_4$ and grew a single crystal. We found no data in the literature on this substance. Our interest in it was due to the fact that $Co^{2+}$ is quite strongly different from $Mn^{2+}$ and $Cu^{2+}$ in a magnetic respect. $Co^{2+}$ has an orbital moment and a large spin–orbit interaction. Incorporating it in a crystalline matrix gives rise to a spin–lattice interaction and strong magnetic anisotropy. Many magnets containing $Co^{2+}$ manifest an Ising character of the interaction with a large value of the anisotropy, and spin-reorientation magnetic transitions are observed in them. As to the type of magnetic anisotropy, many cobalt-containing compounds with a partially frozen orbital moment in the ordered state display easy-plane properties, which should be manifested in the resonance properties under applied pressure. Meanwhile, judging from the data in the iterature [4], many representatives of this system possess the Dzyaloshinskii–Moriya interaction.

Compounds of this class are attracting interest not only because of their uniqueness as two-dimensional magnets but also because of the presence of structural phase transitions in many of them at temperatures of 280 K and above. The forces that bring these transitions about are completely different from those in three-dimensional perovskites. In the case under discussion the structural phase transitions are due primarily to a change of the motion or configuration of the organic chains, and thus they are similar to organic membrane structures [5–7].

The combination of low dimensionality of the magnetic structure and the canting of the magnetic sublattices makes crystals of this family particularly interesting for exploring the resulting features of the magnon energy spectrum and phase transitions.

The goal of this study was to investigate the features of the static magnetic and resonance properties of $(NH_3)_2(CH_2)_3CoCl_4$ and to determine the magnetic parameters of the system.

The substance studied was obtained by the reaction of the metal chloride with propylenediammonium by the scheme

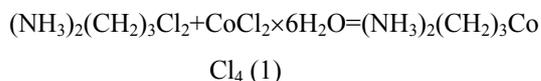
$(NH_3)_2(CH_2)_3Cl_2 + CoCl_2 \times 6H_2O = (NH_3)_2(CH_2)_3CoCl_4$ (1)

Single crystals were grown from solution. The solvent used was water. The saturated solution of the substance obtained (1) was placed in a temperature-controlled chamber with temperature stabilization of ±0.2. The solution was slowly evaporated at T = 40 C. It is known that in the series of compounds $(NH_3)_2(CH_2)_3CoCl_4$ the crystals for even n have a monoclinic structure and those with odd n have an orthorhombic structure [8]. Thus it is assumed that this compound crystallizes in the orthorhombic group $D_{16}^{2h}$ ($P_{nma}$) with z = 4. The structure consists of wrinkled layers of cobalt separated by long organic molecules. The distance between neighboring layers along the b axis is ~8.4Å. The wrinkling is due to the tilting of the $CoCl_6$ complexes by approximately 4 in the direction of the a axis.

For measurement of the magnetization we used a vibrating magnetometer, and the magnetic susceptibility measurements were made by an induction technique. All of the measurements were made in the temperature interval 0.5–50 K and the field interval 0–20 kOe. Temperatures below 1.8 K were reached by pumping on 3He, while fields up to 20 kOe were produced by a superconducting solenoid. Measurements at low fields from 0 to 1000 Oe were made in a copper solenoid. For the resonance studies a wide-band direct-gain EPR spectrometer was used.

## MAGNETIC PROPERTIES

The main preliminary experimental research done on new magnetic materials is to investigate the magnetic characteristics. In this paper we present the most important experimental results on the magnetic susceptibility and magnetization of the compound $(NH_3)_2(CH_2)_3CoCl_4$. Preliminary measurements of the magnetic susceptibility $\chi(T)$ in this compound have permitted us to determine the principal magnetic axes $x$, $y$, $z$, which apparently coincide with the crystallographic axes $a$, $b$, and $c$ (there have been no x-ray studies of this crystal). In neighboring magnetic layers the spins are aligned antiparallel to each other and oriented along the $y$ axis, which is the easy axis of anisotropy. The easy axis is perpendicular or nearly perpendicular to the plane of the layers. In the high-temperature region 5–100 K the magnetic susceptibility is described by the Curie–Weiss law: $\chi(T) = C/(T-\theta)$ ($\theta$ is the Curie temperature). The absence of noticeable anisotropy of the magnetic susceptibility at high temperatures is characteristic of Heisenberg magnets. With decreasing temperature, $\theta$ increases monotonically, reaches a maximum at T ≈ 2 K, and then decreases. The $\chi(T)$ curve along the $x$ axis has a large peak with a maximum at around 2 K. This peak exists in a narrow interval of temperatures and is easily suppressed by a small magnetic field. Therefore the magnetic susceptibility was measured above 4.2 K by a vibrating magnetometer in a field of around 1 kOe, and below 4.2 K by an inductive bridge in an ac field below 1 Oe.

Figure 1 shows the temperature dependence of the magnetic susceptibility at a frequency of 1 kHz in the temperature interval 1.5–5 K. The measuring ac field was equal to 0.18 Oe. As is seen in Fig. 1, at a temperature of 2.05 K susceptibility maxima are observed along all three magnetic axes, the component along the $x$ axis jumping sharply by almost two orders of magnitude in comparison with the magnetic susceptibility at helium temperature. Near these anomalies the behavior of χ is influenced noticeably by the strength of the magnetic field. Below 2 K the magnetic susceptibility increases sharply. We attribute such behavior of χ(T) to antiferromagnetic ordering and to the formation, in the ordered phase of the crystal, of a spontaneous ferromagnetic moment lying in the *xz* plane.

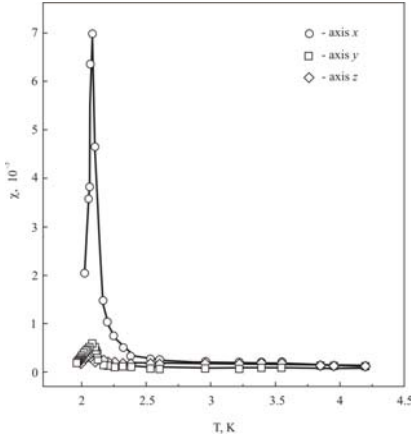

FIG. 1. Temperature dependence of the magnetic susceptibility of the $(NH_3)_2(CH_2)_3CoCl_4$ single crystal along the magnetic axes x, y, z.

To check the correctness of the interpretation of the experimental results for χ(T) in the vicinity of 2 K we have made additional measurements of the real and imaginary parts of the magnetic susceptibility χ' and χ'' of the $(NH_3)_2(CH_2)_3CoCl_4$ crystal along the *x* axis as a function of temperature at frequencies of 30, 300, and 3000 Hz. The results of the measurements at 3000 Hz are shown in Fig. 2. The experiment quite reliably reveals growth of χ' and the presence of significant absorption in the vicinity of 2 K. This attests to a relaxation character of the magnetic losses in this crystal, apparently due to the appearance of domains and to the motion of domain walls. The temperature hysteresis is plainly visible. The character of the curves and the relationship between χ' and χ'' depend on the frequency. As the frequency is lowered, χ' tends toward the static susceptibility and χ'' tends toward zero. The behavior shown is typical of the temperature dependence of χ in the presence of magnetic domains.

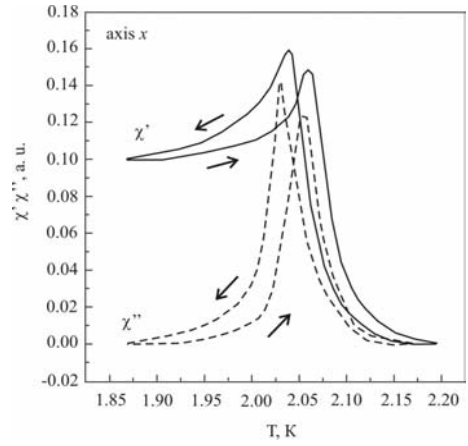

FIG. 2. Temperature dependence of χ'(T) and χ''(T) at a frequency of 3000 Hz, H||x.

The temperature dependence of the inverse susceptibility $\chi^{-1}(T)$ in the temperature interval 4.2–50 K along the three magnetic directions is shown in Fig. 3. The curves are only slightly different and have a paramagnetic character (the $\chi^{-1}(T)$ curves are close to being straight lines). From the data shown, the following values of the Curie temperature θ are determined: $\theta_z$ = -4.2 K, $\theta_x$ = -1.8 K, $\theta_y$ = -5.5 K; these values are indicative of an antiferromagnetic interaction between the $Co^{2+}$ ions in this compound.

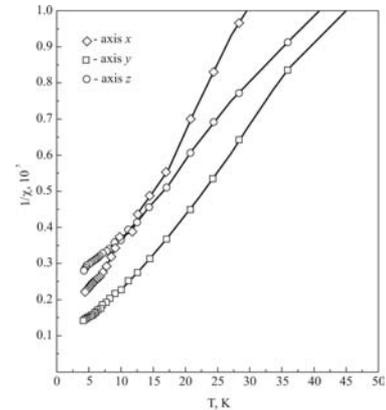

FIG. 3. Temperature dependence of the inverse magnetic susceptibility 1/χ along the magnetic axes x, y, z.

The temperature dependence of the magnetic susceptibility, χ(T), χ'(T), and χ''(T), was supplemented by measurements of the field and temperature dependences of the magnetization, M(T) and M(H). Figure 4 shows M(T) along the *x* axis in magnetic fields of around 1 and 10 Oe, produced by a copper solenoid. The sharp rise of the magnetization near $T_N$ is evidence of the onset of a spontaneous magnetic moment. Another distinctive feature of M(T) is the presence of temperature hysteresis in this substance at the transition to the ordered state at T =2.05 K. The value of the hysteresis was around 0.02 K.

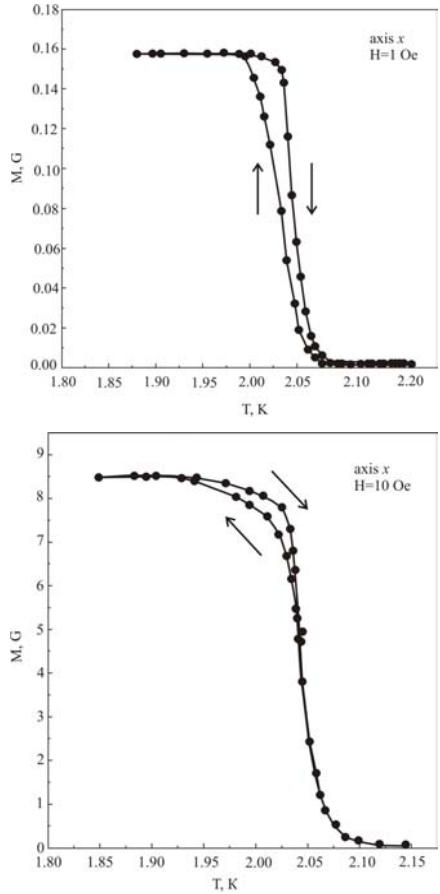

FIG. 4. M(T) curves in external magnetic fields of 1 Oe (a) and 10 Oe (b), H||x.

Thus the temperature curves of M(T) and $\chi'(T)$, $\chi''(T)$ measured along the $x$ axis in two independent experiments demonstrate the presence of temperature hysteresis at the transition to the ordered state and attest that the magnetic ordering in this substance occurs as a first-order phase transition. Usually the transition to a magnetically ordered state occurs as a second-order magnetic phase transition. In our opinion, the magnetic ordering is accompanied by a first-order phase transition because of strong magnetoelastic interaction.

In spite of the fact that the preceding measurements of the temperature dependence M(T) and $\chi'(T)$, $\chi''(T)$ attest to the presence of a spontaneous magnetic moment and temperature hysteresis along the $x$ axis for T <$T_N$ and, consequently, the presence of magnetic domains with magnetization vector along this direction, the M(H) curves did not exhibit hysteresis in magnetic fields up to 40 Oe (Fig. 5). In those experiments the resolution with respect to field was of the order of ±0.5 Oe. Consequently, the width of the hysteresis loop, if it exists, is much less than 1 Oe.

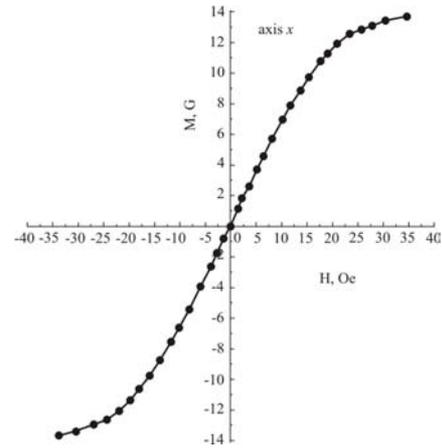

FIG. 5. Magnetic-field dependence M(H) along the $x$ axis in magnetic fields to 40 Oe.

To trace the transformation of the field dependence of the magnetization M(H) along the $x$ axis at the transition through the ordering temperature, we made measurements of the magnetization in fields up to 11 kOe at several temperatures from 4.2 K to 1.8 K. The results are shown in Fig. 6. It is interesting to note that the slope of M(H) in fields above 50 Oe is practically independent of temperature and field within the given limits of field and temperature. Along directions other than $x$ there is no spontaneous magnetic moment in the ordered state.

The general picture of the magnetization curves in fields up to 20 kOe along the three magnetic axes at temperatures of 4.2 and 0.5 K can be seen in Fig. 7. At helium temperature the M(H) curves along the three axes are straight lines (a paramagnetic state) in the investigated range of applied field. The slopes of these lines (the susceptibilities) are not very different. At 0.5 K the picture is radically altered. While the M(H) curve along the $z$ axis remains linear, the curves along the $y$ and $x$ axes differ sharply not only from the curve along the $z$ axis but also from each other. Along the $y$ axis, the slope of M(H) at low fields is almost an order of magnitude less than at helium temperature. Then, at a field of 11 kOe, the slope increases rather abruptly, reaching values two orders of magnitude larger than the initial values. When the magnetization becomes equal to 50 G, the slope decreases to a value of the order of the value of the slope in the paramagnetic state and maintains that value to the highest fields reached in the experiments, ≈20 kOe. Along the *x* axis, at 0.5 K the M(H) curve has approximately the same form as it takes on at the ordering temperature, namely, sharp growth to 14 G at very low fields, then moderate growth to the maximum field, with a slope close to that in the paramagnetic state. From the shape of these curves one can conclude that the magnetization vectors lie in the *xy* plane: along the *y* axis an antiferromagnetic character is observed, with the formation of a spin-flop phase in a field of 11 kOe, and along the *x* axis a spontaneous magnetic moment appears at $T_N$, with the formation of magnetic domains with magnetization parallel and antiparallel to the *x* axis.

The change of M(H) in the vicinity of the spin-flop transition occurs rather smoothly, so that it can be stated that we are dealing with a first-order phase transition. The smearing of the transition may be caused by the poor quality of the samples, paramagnetic impurities, imprecision of the mounting of the sample, etc. We note, however, that among the high-quality samples in this series of manganese and copper compounds, $(NH_3)_2(CH_2)_3MnCl_4$ is the only one in which a sharp spin-flop transition is observed. Therefore the elucidation of the physical causes of such a feature is a problem for further study.

The appearance of a spontaneous magnetic moment in antiferromagnets, as a rule, is a consequence of noncollinearity of the magnetic sublattices. The mechanisms that lead to canting of the spins include single-ion anisotropy, antisymmetric superexchange, and anisotropy of the *g* factor. In the case of $Co^{2+}$ apparently all of these mechanisms can give rise to weak ferromagnetism. As is shown in the paper by Turov [9] the presence of antiferromagnetic ordering ($\vec{l} \neq 0$), by virtue of the symmetry properties of the crystal, should give rise to a spontaneous magnetic moment ($\vec{m} \neq 0$), only in the case when the equation for the magnetic energy contains terms of the form $\vec{m}_\alpha \vec{l}_\beta$ (α, β = x, y, z) or terms of higher order in $\vec{l}$ but linear in $\vec{m}$. The space group of the crystal is $P_{nma}$, which, according to theory, admits the existence of weak ferromagnetism.

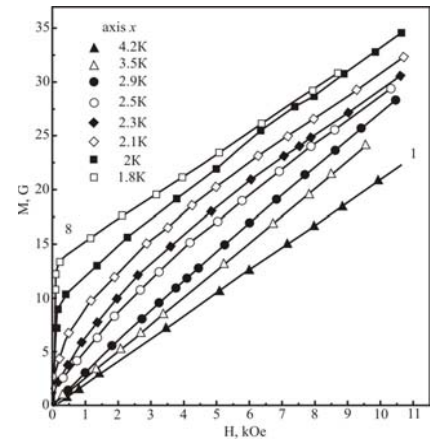

*FIG. 6. Isothermal curves of M(H) in magnetic fields to 11 kOe, H||x.*

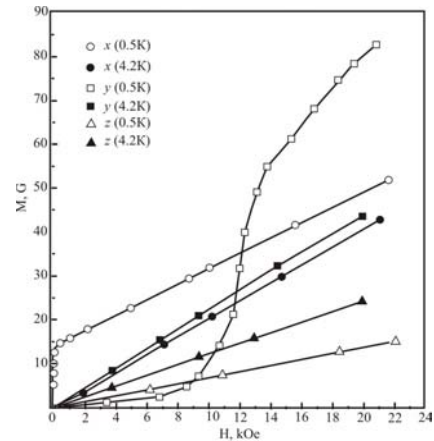

*FIG. 7. M(H) curves in fields to 20 kOe along three magnetic axes at fixed temperatures of 0.5 and 4.2 K.*

Using the experimental curves of M(H), one can obtain quantitative values of the exchange interaction field He and sublattice canting angle. From the slope of the M(H) curve after the spin-flop transition we determined the saturation magnetization in this compound, $2M_0$ = 150 G, and the value of the exchange interaction field, $H_e$ = 46 kOe. Knowing the values of the saturation magnetization $M_0$ = 75 G and the spontaneous moment $M_s$ = 14 G, one can estimate the purely geometric canting angle of the magnetic sublattices: $\gamma$ = arcsin($M_s/2M_0$) ≈ 5.3. All of the quantitative values of the physical quantities are obtained with an error of around 5%.

Measurements of M(H) in the vicinity of the spin-flop transition field, equal to 11 kOe at T =0.5 K, permits estimation of the value of the anisotropy field $H_a$ according to the formula [4]

$$H_a = \frac{H_{sf}^2 (\frac{1-\chi_\parallel}{\chi_\perp})}{2H_e} \quad (2)$$

where $(\frac{1-\chi_\parallel}{\chi_\perp})$ is a quantity of the order of unity, and then $H_a$ =1.3 kOe.

We then find the anisotropy parameter $\alpha$ as the ratio of the anisotropy field $H_a$ to the exchange field $H_e$: $\alpha = \frac{H_a}{H_e}$ ~ 2.8 × 10$^{-2}$, which is not a very small quantity.

The results of the measurements of $\chi$(T) and M(H) show that the compound $(NH_3)_2(CH_2)_3CoCl_4$ has a canted antiferromagnetic structure. The resultant moment along the *x* axis is established unambiguously. The spin structure is characterized by a canted antiparallel arrangement of spins away from the easy axis *y* toward the *x* axis, in agreement with the crystal structure, in which the $Co^{2+}$ octahedra are inclined to the *xy* plane.

RESONANCE STUDIES

As is well known, resonance methods of investigating magnetically concentrated crystals can give voluminous information about the energy spectrum, interactions, and the structure of the magnetic subsystem and permit investigation of the frequency–field diagrams of the AFMR spectra, since from the character of the frequency–field curves of the AFMR one can assess the type of magnetic structure of a substance and obtain direct information about the values of the isotropic and anisotropic interactions of the magnetic ions.

We studied the resonance properties of the $(NH_3)_2(CH_2)_3CoCl_4$ single crystal in the frequency range 40–120 GHz at a temperature of 1.7 K. For using the wide-band spectrometer the samples were placed in a resonant cavity of the corresponding frequency range, which had provisions for rotation of the sample within the cavity in a vertical plane. The absorption signal was detected by a reflex-type radio spectrometer. For control of the temperature we used an attachment placed in the superconducting solenoid (producing a magnetic field of 75 kOe) which permitted experiments in the temperature interval 1.7–100 K.

As we have shown above, in the crystal under study the magnetic ordering temperature $T_N$ =2.05 K. The resonance measurements were made at 1.7 K. The closeness of the working temperature to $T_N$ causes certain difficulties in the AFMR studies, since the absorption lines are broadened.

The AFMR absorption spectra were measured as follows.

First the EPR spectra along the principal magnetic axes were measured at a frequency of 78.2 GHz and a temperature of 4.2 K. The following values of the effective g factors were obtained: $g_z$ =1.93, $g_y$ =3.34, $g_x$ =3.15. Of course, these values of the g factors at helium temperature are apparently altered by the exchange interaction, since the measurement was made near the Néel point.

After the temperature was lowered to the minimum temperature of the experiment, 1.7 K, the frequency–field curves of the AFMR spectrum of the $(NH_3)_2(CH_2)_3CoCl_4$ single crystal in the *xy* plane were investigated for a perpendicular polarization of the microwave field ($\vec{h} \perp \vec{H}$). The results are presented in Fig. 8.

The Hamiltonian of an orthorhombic antiferromagnet (AFM) in an external magnetic field has the form [9]

$$H = -\frac{\delta l^2}{2} + \frac{\beta_1 l_x^2}{2} + \frac{\beta_2 l_y^2}{2} + d_1 m_x l_y + \tau_1 H_x l_y - \tau_2 H_y l_x - \vec{m}\vec{H} \quad (3)$$

where δ is the exchange constant, $\beta_1$ and $\beta_2$ are the anisotropy constants, $\vec{m}$ is the ferro magnetic vector, and $\tau_1$ and $\tau_2$ are quantities characterizing the anisotropy of the g factor. The magnetic states of this Hamiltonian are analyzed thoroughly in Ref. 10.

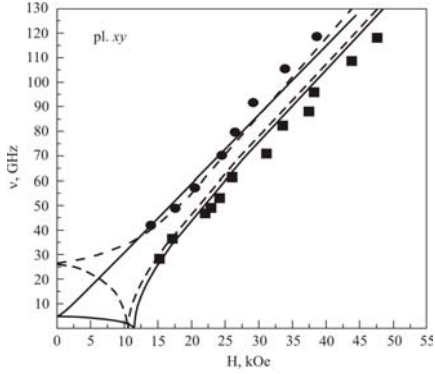

FIG. 8. Frequency–field curves of the AFMR in $(NH_3)_2(CH_2)_3CoCl_4$ for H||y and H||x, T =1.7 K. The dot-and-dash line shows the proposed frequency–field dependence for T =0.5 K.

When the external magnetic field is oriented along the easy axis y one observes an AFMR that we attribute to the excitation of a spin-flop mode in fields exceeding the tipping field of the magnetic sublattices. The value of the transition field $H_t$ = 12.4 ± 0.1 kOe. It should be noted that the experimental dependence is somewhat different from the theoretical curve of Ref. 10, which is described by the expression

$$\nu = \sqrt{H^2 - H_{sf}^2} \quad (4)$$

The quantitative agreement with experiment is clearly unsatisfactory.

The disagreement can be explained by the influence of the demagnetizing factors on the resonance of the tipped phase at H>$H_t$, the inclusion of which in the analysis will lead to renormalization of the parameters of the spectrum. The disparity between the classical dependence of the spin-flop mode on the value of the magnetic field can apparently also be attributed to the value of the anisotropy between the easy and intermediate directions of the antiferromagnetic vector in the system rather than to the specifics of the low dimensionality. We note that we did not observe a high-frequency gap along the easy axis in $(NH_3)_2(CH_2)_3CoCl_4$ in the frequency range investigated.

We have investigated the frequency–field dependence of the AFMR described by the following expression [10] when the external magnetic field is oriented along the intermediate axis x:

$$\nu = \sqrt{2H_e H_a + H^2 + HH_d} \quad (5)$$

which takes into account the Dzyaloshinskii interaction and the value $H_d$ = 3 kOe. In the expression for the resonance frequency we have not taken into account the anisotropy of the g factor of the $Co^{2+}$ ions, which is manifested for different orientations of the magnetic field, since the $g_x$ and $g_y$ factors are practically equal. Extrapolation of the frequency–field dependence at T =1.7 K gives a gap value of 5.1±0.3 GHz at zero magnetic field. For lowdimensional systems the temperature behavior of the gaps is determined in a wide range of temperatures by the interaction of magnons and coincides with the temperature dependence of the sublattice magnetizations at zero magnetic field and agrees with the Oguchi–Honma spin-wave theory [11].

Unfortunately, the value of the exchange interaction field determined from the static magnetic measurements with an error of up to 5% and also the significant AFMR linewidth do not permit recovery of the shape of the spectral branches to good enough accuracy.

The set of resonance and magnetic measurements allows us to recover the magnetic structure of the compound under study and to determine the magnetic parameters describing the static and high-frequency properties of $(NH_3)_2(CH_2)_3CoCl_4$. A fragment of the proposed magnetic cell of $(NH_3)_2(CH_2)_3CoCl_4$ and the configuration of the magnetic moments in the layers are shown in Fig. 9.

The minimum energy of excitation of magnons and the effective magnetic parameters at a temperature of 1.7 K have the following values: $\nu_1$ =5.1 GHz, $H_{sf}$ = 12.4 kOe, $2H_e$ = 92 kOe, $H_{a1}$ =0.035 kOe, $H_d$ = 3 kOe.

Here the effective magnetic fields $H_{a1}$, $H_d$, and $2H_e$ correspond to the constants $\gamma_1$, $d_1$, and $\delta$ of the Hamiltonian (3). The values of the g factors obtained in the EPR experiments and used in the description of the magnetically ordered state are as follows:

$$g_z = 1.93; \quad g_y = 3.34, \quad g_x = 3.15$$

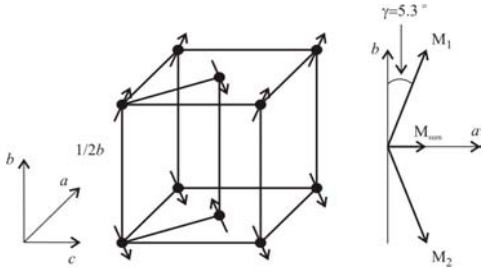

FIG. 9. *Orientation of the moments of the sublattices in the magnetic cell of $(NH_3)_2(CH_2)_3CoCl_4$ and their mutual arrangement in neighboring layers.*

However, the large difference of the gaps and, hence, the large difference of the values of the effective magnetic anisotropy fields allow us to assume that the dominant anisotropy in the compound $(NH_3)_2(CH_2)_3CoCl_4$ is apparently of the easy-plane type, with a weak anisotropy within the plane. This is not a great rarity among cobalt-containing compounds, for which the single-ion anisotropy can exceed the exchange interaction, in which case the oscillations of the antiferromagnetic structure will correspond to easy-plane. These discussions are consistent with static magnetic studies, which indicate that the magnetic moments lie in the *xy* plane.

Thus we have obtained the following main experimental results, which in our opinion are decisive.

We have investigated the temperature dependence of the magnetic susceptibility and magnetization of the $(NH_3)_2(CH_2)_3CoCl_4$ single crystal along the principal axes. We have determined the magnetic ordering temperature $T_N = 2.05$ K and the spontaneous magnetic moment along the *x* axis. We have shown that the phase transition to the antiferromagnetic state occurs as a first-order transition. We have determined the paramagnetic Curie points $\theta$.

We have investigated the isothermal M(H) curves in fields to 20 kOe. At T = 0.5 K we have observed a sharp increase of the magnetization in a field of 11 kOe along the *y* axis due to the spin-flop transition. The value of the exchange interaction in the given crystal is $H_e = 46$ kOe. According to the experimental data we have constructed the magnetic structure of the crystal.

A study of the frequency–field dependence of the AFMR spectrum in $(NH_3)_2(CH_2)_3CoCl_4$ (T = 1.7 K) has revealed a low-frequency gap in the spin-wave spectrum and has yielded the value of the effective magnetic anisotropy field leading to the gap of the low-frequency mode.